\documentclass{article} % For LaTeX2e
\usepackage{colm2024_conference}

\usepackage{microtype}
\usepackage{hyperref}
\usepackage{url}
\usepackage{booktabs}
\usepackage{epigraph}

\title{And Then the Hammer Broke: Seeing Machine Vision \\ \vspace{-3mm} \\ \large \normalfont \textit{Reflections on Machine Ethics from Feminist Philosophy of Science}}

\author{Andre Ye \\
Department of Philosophy \\
Paul G. Allen School of Computer Science \\
University of Washington \\
\texttt{andreye@uw.edu}
}

\usepackage[notes,backend=biber]{biblatex-chicago}
\bibliography{bib}

\usepackage{listings}
\usepackage{graphicx}

\colmfinalcopy % Uncomment for camera-ready version, but NOT for submission.
\begin{document}

\maketitle

\begin{abstract}
Vision is an important metaphor in ethical and political questions of knowledge. The feminist philosopher Donna Haraway points out the ``perverse'' nature of an intrusive, alienating, all-seeing vision (to which we might cry out ``stop looking at me!''), but also encourages us to embrace the embodied nature of sight and its promises for genuinely situated knowledge. Current technologies of machine vision -- surveillance cameras, drones (for war or recreation), iPhone cameras -- are usually construed as instances of the former rather than the latter, and for good reasons. However, although in no way attempting to diminish the real suffering these technologies have brought about in the world, I make the case for understanding technologies of computer vision as material instances of embodied seeing and situated knowing. Furthermore, borrowing from Iris Murdoch's concept of moral vision, I suggest that these technologies direct our labor towards self-reflection in ethically significant ways. My approach draws upon paradigms in computer vision research, phenomenology, and feminist epistemology. Ultimately, this essay is an argument for directing more philosophical attention from merely criticizing technologies of vision as ethically deficient towards embracing them as complex, methodologically and epistemologically important objects.
\vspace{2mm}

\end{abstract}

\epigraph{The eye sees only what the mind is prepared to comprehend.}{Henri Bergson}

\section{The Hammer}

Technologies of machine vision deeply unsettle us: behind the dark film of the camera lens lies some perverted nonhuman Other who records, surveils, and sees everything. ``Vision in this technological feast become unregulated gluttony,'' Donna Haraway writes in her seminal essay ``Situated Knowledges''.\autocite[581]{Haraway1988}.
``Seeing everything from nowhere... this eye fucks the world to make techno-monsters.'''
Indeed, our world is thoroughly populated with stories of these techno-monsters. 
Facial recognition systems used by the New York State Department of Corrections incorrectly classify inmates' family members as barred persons, especially for Black and Brown people.\footnote{\lstinline{https://www.nyclu.org/en/news/inaccurate-facial-recognition-prisons-keeping-families-apart}}
The Chinese state runs computer vision algorithms on data pools from an extensive system of surveillance cameras to track individuals at shockingly granular levels. These systems not only inform authorities of possible risk, but also on how to confront it: for instance, the Hikvision system used in Tianjin even produces detailed psychological profiles of surveilled subjects.\footnote{\lstinline{https://www.nytimes.com/2022/06/25/technology/china-surveillance-police.html}}
War drones directed by computer vision algorithms are now capable of pursuing and eliminating targets with higher precision and speed despite reduced human direction. 
In 2020, the Turkish-manufactured lethal drone Kargu was used to autonomously target retreating soldiers allied with Libyan General Khalifa Haftar -- speculated to be the first instance of such a wartime machine-led killing.\autocite{UNSC2021}
But the technologies at the core of these techno-monsters are the same as the ones we find in the phones we blankly stare at to unlock,  the gender-switching and dog-mask filters on Instagram and Snapchat, and the virtual meeting software that blurs our backgrounds and `touches up' our virtual appearances. 
At both the most mundane and the most horrific levels of life, we find ourselves in a world dreadfully saturated with seeing machines.

Sick of being the perpetual object of this perverse sight, we develop critical accounts of the seeing function of seeing machines: we demonstrate how it can represent an oppressive force,\footnote{See: R. A. Waelen, ``The ethics of computer vision: an overview in terms of power'' (\textit{AI Ethics}, 2023).} how it reifies existing instances of structural inequality,\footnote{See: Timnit Gebru, "Race and gender" from \textit{The Oxford Handbook of Ethics of AI} (Oxford, 2020).} how it damages human life across mental, social, political, and economic dimensions.\footnote{See: Paul Scharre, \textit{Army of None: Autonomous Weapons and the Future of War} (W. W. Norton $\&$ Company, 2019).}
In doing so, we accuse machine vision of playing something like a `god trick': in Haraway's words, the illusion ``of seeing everything from nowhere''\autocite[581]{Haraway1988} and of ``infinite vision.''\autocite[582]{Haraway1988}
\textit{Deus ex machina}: a ``scientific and technological, late-industrial, militarized, racist, and male-dominant''\autocite[581]{Haraway1988} god arises from the machine, and imposes the will of its commanders -- capitalists, corporate leaders, politicians -- ceaselessly and totally across the earth.

But perhaps we should (carefully) ask if we have not also played a role in constituting this god trick.
We have criticized the seeing function of machine vision as it is exercised in mature form, but have we asked yet about how this seeing function comes to be: about its conditions for emergence, and about the labor involved in making it emerge?
Machine vision has seen us (all too much); have we endeavored in turn, to really see it?
Consider Heidegger's hammer: hammer in hand, the carpenter goes to work driving the nails into the wood, ceasing to become conscious that it is a separate object at all.
When the hammer functions perfectly, does the carpenter not acquire the quality of god, to shape the object to her will?
To play god requires technologies that function entirely ideally, without resistance, that allow for the exercise of will upon the world.
But Heidegger's hammer is not Thor's divine hammer, Mjölnir: Heidegger's hammer breaks, and the carpenter becomes aware of the hammer as a particular object with particular properties, and of herself as a particular being actualizing her particular will with this particular object.
When the hammer breaks, the carpenter really sees the hammer. The god role-play is interrupted.
In criticizing machine vision, have we hastily constituted it as Mjölnir, a perfect instrument for its wielders, and in doing so missed an important part of the picture?
Insofar as machine vision is a material technology, subsisting in computer hardware, photoreceptive sensors, and algorithms written out by warm fingers on cold keyboards in code: is it not instead a Heideggerian hammer?
 
In this essay, I endeavor to answer both of these questions affirmatively by advancing a materially informed reading of machine vision drawn from work both in $\{$cybernetics, computer vision, machine learning$\}$ and $\{$feminist philosophy of science, moral philosophy$\}$.
By recognizing machine vision as a Heideggerian hammer, I claim, we can really see machine vision, just as the carpenter learned to really see the hammer.
This allows us to more critically and substantively take up Haraway's suggestion that ``understanding how these visual systems work, technically, socially, and psychically, ought to be a way of embodying feminist objectivity.''\autocite[583]{Haraway1988}.
My goal in this paper is not to `redeem' machine vision in any moral sense, nor to trivialize critical accounts of it.
Rather, I see myself as contributing to a further, more nuanced understanding of machine vision.
As Haraway declared succinctly in the opening to \textit{A Cyborg Manifesto}, ``blasphemy is not apostasy'':\autocite[5]{Haraway2016} to challenge and prod this body of thought is not to abandon it altogether.
The substance of my argument is twofold.
First, under a close reading of Haraway's illustrations of the god trick in ``Situated Knowledges'', I will argue that while humans play god in Haraway's examples of machine vision, in the context of contemporary machine vision, the machines play god -- but in a materially subversive way that resists the god trick.
Secondly, borrowing from Iris Murdoch's notion of `moral vision', I suggest that -- even if we cannot ascribe a particular ethical orientation to it -- machine vision makes its internal perceptual and cognitive workings transparent and therefore directs our labor towards seeing moral reality.

\section{``God is dead, and it has killed itself''}

Looking at those colorful photos of vast outer planets hundreds of millions of miles away captured by giant telescopes, one might begin to believe that really are seeing the true majesty of the celestial bodies. These images, Haraway writes, appear to ``let the viewer `experience' the moment of discovery in immediate vision of the `object'.''\autocite[582]{Haraway1988}
Likewise, imaging devices capture the microscopic world at wavelengths imperceivable to humans, opening up a mysterious microbiological world of ``defending T cells and invading viruses.''\autocite[582]{Haraway1988}
Wielding Thor's hammer, we cease to become aware of the instruments of vision and thereby buy into the god trick: both that we can see everything in the universe as they really are, and that this view is from `nowhere', in that my own vision may be subjectively embodied but my instruments of vision allow me to access objective vision `beyond' my body. 
Importantly, the instruments of vision are not the `source' or `constitutor' of the god trick.
A digital telescope is `merely' a technological object which records radiation from space, and microscopic imaging techniques `merely' ping organic structures with radiation to collect data about their density and organization.
That is to say: they are material machines.
It is us humans, like the carpenter, who use the outputs of these machines to generate fantasies of knowledge -- admittedly beautiful fantasies -- of the cosmos' grand beauty, and of the countless miniature wars fought within the very matter of our body.
This is not only observed in the scientific domain, but in our world of media.
Scrolling through infinite video feeds on the Internet, one may subscribe to the illusion that they are seeing the world as it is, ceasing consciousness of the army of cameras -- pointed in particular directions by particular people -- which produced each of those videos.
Haraway understands that these instruments are Heideggerian hammers, that they break: ``The only people who end up actually believing and, goddess forbid, acting on the ideological doctrines of disembodied scientific objectivity... are nonscientists.''\autocite[576]{Haraway1988}
The astronomer who runs the computer software which converts raw radiation signals into ``coffee table color photographs''\autocite[583]{Haraway1988} is not under the illusion that we are looking at the cosmos when we peer into those images, nor that the fantasies which spring from those images are imposable back onto the `real' universe. Haraway is right, then, when she notes that scientists ``might be our allies.''\autocite[594]{Haraway1988}

But I worry that the examples raised in this 1988 text of Haraway's, ``Situated Knowledges'', may not itself be sufficiently situated in our present (scientific and technological) knowledge.
Machine vision, I argue, does not participate in the god trick in the same way anymore.
Haraway's machines -- the microscopes and telescopes -- are content with being instruments wielded by humans.
But contemporary machine vision aspires for much more. 
We can get a sense for this shift by understanding how the researchers that built our contemporary technologies of machine vision understand (machine) vision.
Li Fei-Fei and Ranjay Krishna, two prominent researchers in the field, take vision to be ``the ability to see'' and as ``central to intelligence''.\autocite[86]{FeiFei2022}
Central to their analysis is that ``computer vision, like human vision, is not just perception; it is deeply cognitive.''\autocite[95]{FeiFei2022}
The first monumental goal for computer vision as a budding research area in the late 20\textsuperscript{th} century was the recognition of particular objects in a visual field, such as a house or a dog.
This is a trivial task for most humans, but for a computer, an image is merely a collection of numbers arranged along height, width, and color axes: so, from an algorithmic perspective, there is no immediately clear association of pixels which indicates the presence of a house, a dog, or some other category.
To adapt, Fei-Fei and Krishna write, machine vision research shifted its focus from algorithms to data.
Researchers had initially hand-designed algorithms to check for particular patterns and important features in the image, but these failed to capture the variability and dynamic nature of object categories.
Machine vision started the ``search for a new approach with one key assumption: even the best algorithm would not generalize well if the data it learned from did not reflect the real world.''\autocite[87]{FeiFei2022}
This is a significant paradigm shift: building machine vision is no longer principally about building a better machine, but rather about building a better map of the world.
This map, Fei-Fei and Krishna suggest, must achieve three design goals: ``scale (a large quantity of data), diversity (a rich variety of objects), and quality (accurately labeled objects).''\autocite[88]{FeiFei2022}
The instances of machine vision that we encounter in our current world -- the New York facial recognition model, Chinese psychological profiling systems, autonomous war drones, but also social media filters and virtual meeting software -- are farther from Haraway's telescopes and closer to this new kind of machine vision.

\begin{figure}[t]
    \centering
    \includegraphics[width=\textwidth]{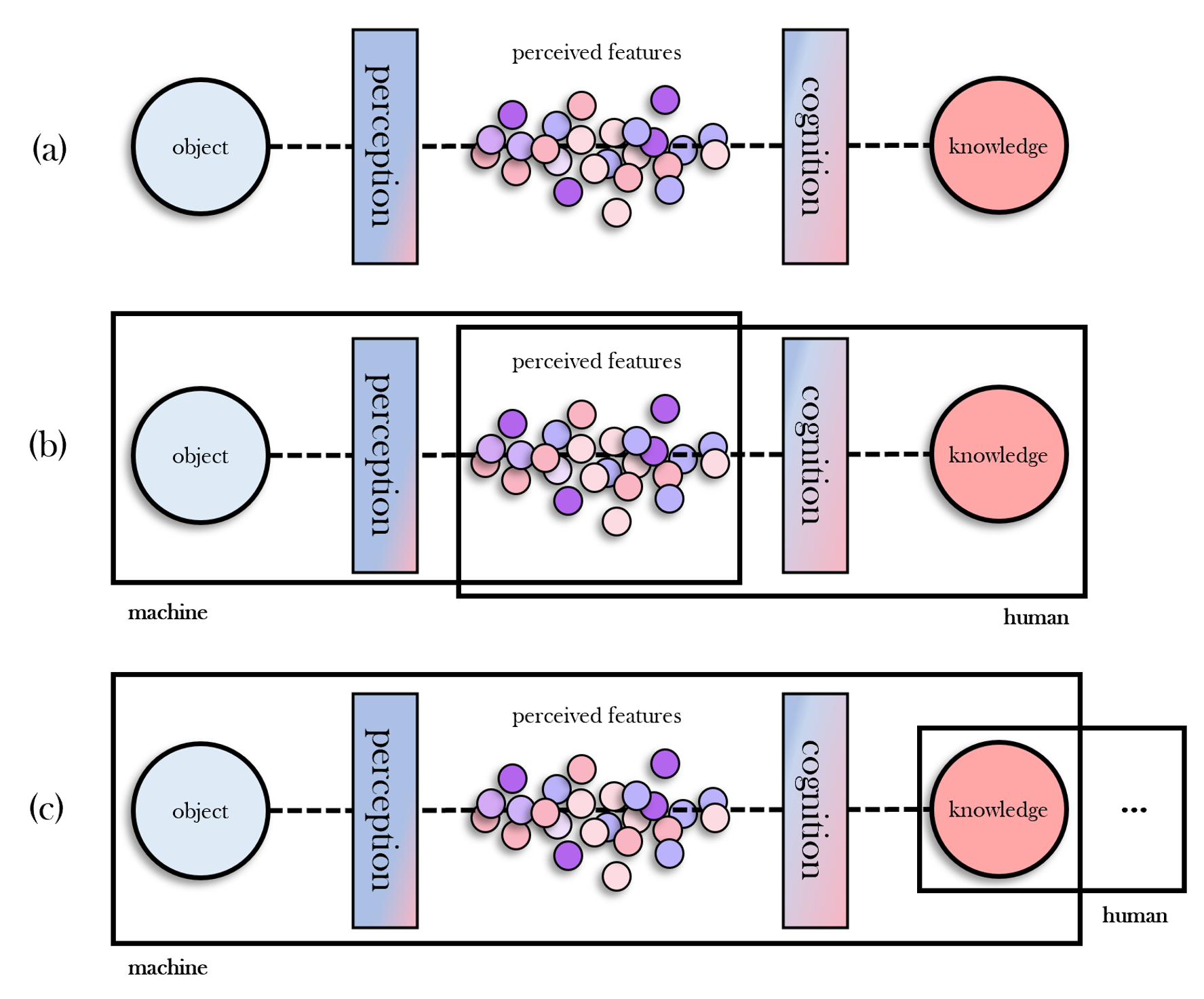}
    \caption{a) \textit{Vision in general}: objects are mapped to perceived features, which are cognized as knowledge. b) \textit{Vision for Haraway’s machine vision}: the machine perceives the object and generates perceived features, which we cognize into (false) knowledge. c) \textit{Vision for our contemporary machine vision}: the machine both perceives and cognizes the object, and we take the resulting knowledge and use it for some other purpose.}
    \label{fig:enter-label}
\end{figure}

To understand exactly what kind of god trick is played by the new machine vision, we need to clarify what is meant by `vision'.
We can begin with Fei-Fei and Krishna's suggestion that vision and sight are not only perceptive but also cognitive.
We not only see objects and places, but also perspectives, arguments, emotions -- those complex things to which we respond, ``I see.''
Seeing an object begins with perception: the task of perception is to map the object of vision to a set of salient perceived features.
For instance, the Gestalt school of psychology posits that humans perceive visual information through a set of basic structures, such as proximity (objects close together are grouped as one) or closure (we can `impose' or `close' shapes onto broken or noisy visual information which resembles it).\autocite{Wagemans2012}
The perceived features of a seen object can also be empirically measured by identifying which parts of the brain activate due to visual stimulus.\autocite{Epstein2019}
In his pioneering phenomenological method, Husserl refers to these features as ``sense-data''.\autocite[39]{Husserl1960}
Suppose you sit before a table, Husserl suggests.
The immediately perceived data includes the surface of the table and possibly two of the legs.
However, we do not see merely a surface and two legs but a table, a three-dimensional body with three or four legs.
This act of \textit{cognition} produces knowledge of the seen object, which Husserl calls \textit{cogitata}.
It may not only be the table as a unitary physical object, but also qualities such as its beauty or invitation to action (to place things upon it, to write upon it) may become apparent to us.
Note that the temporal precedence of perception before cognition does not mean that cognition does not act upon perception: indeed, the very visual features that we immediately perceive on may be conditioned by cognitive beliefs.\footnote{See Benjamin Whorf Lee, “Language, mind, and reality” (\textit{Epistemology and Philosophy of Science}, 2016).}
Nevertheless, such an understanding of vision as perception-cognition is shared across computer vision, artificial intelligence, psychology, neuroscience, and phenomenological approaches to philosophy.
In Haraway's examples of machine vision, the machine takes the role of perception, while the human takes the role of cognition.
Telescopes collect ``sense-data'' from the planets and present them to humans, which might cognize them into knowledge as varying as the beauty of the cosmos to scientific observations about particular astronomical bodies.
But in the contemporary machine vision that Fei-Fei and Krishna describe, the machine is responsible \textit{both} for perception and cognition: humans do not need to cognize from perceived features but rather accept the produced knowledge for other uses (the knowledge may be denied, verified, integrated, etc.). 
To carry on with the example of astronomy: deep learning models directly observe observations from an astronomical field, determine the optimal way to `perceive' the most informative features, and cognize knowledge of what planets are in the field.\footnote{For examples, see: J. Pasquet, “Deep learning approach for classifying, detecting, and predicting photometric redshifts of quasars in the Sloan Digital Sky Survey stripe 82” (\textit{Astronomy and Astrophysics}, 2018). M. Huertas-Company and F. Lanusse, “The Dawes Review 10: The impact of deep learning for the analysis of galaxy surveys” (\textit{Astronomical Society of Australia}, 2022).}
Astronomers do not use these models to `zoom in' on or clarify an object they cannot perceive with their naked eye, but rather to directly produce knowledge of the seen object.
This is a direct result of the shift from machines (producing `better' perceived features) to maps (better replicating the knowledge produced from seen objects).

For Haraway's machine vision, the god trick occurs when we, as cognizers of knowledge, forget that perceived features produced by instruments are perceived, rather than of the direct and full object itself.
\textit{Perceived} means that sense-data is collected in a particular direction by sense-organs particularly oriented in space, just as in Husserl's table -- we may perceive a surface and two legs, but someone lying on the ground but see the bottom and four legs -- or Heidegger's hammer -- the carpenter is a particular being using a particular tool with particular limitations.
This means that the knowledge produced by sight under the god trick is \textit{false} because it claims to reference the original object itself when it has no such claim.
For contemporary machine vision, humans are not given the task of cognizing: models perceive and cognize together.
The god trick, then, cannot be about humans cognizing false knowledge.
Insofar as the god trick is ``seeing everything from nowhere'', the mistake would be to identify knowledge produced by the model as objective.
This is an important difference: in the first case, the trick is that we are playing gods, cognizing universal knowledge with Mjölnir-like instruments at hand; in the second case, the trick is that we think that the models are gods, that the knowledge they produce is universal.

As machine vision has advanced, it has moved -- with respect to the god trick -- from an instrument aiding our cognizing to a god which cognizes without us.
We, in turn, have moved from playing gods to playing the disciples of machinic gods, receiving knowledge passed from upon high and passing it along.
This may seem like a pedantic point: after all, either way, some god trick is being played, and false knowledge is being produced and proliferated.
To the contrary, I argue that that this advancement introduces cracks into the god trick.
Playing god is a mode of belief that humans may entertain, but which is not materially actualizable. 
Insofar as machines are material beings, t\textit{hey are bad at playing god}.
They expose themselves as what they are -- not gods, but carpenters.
Moreover, humans, no longer positioned as playing the god, and therefore detached from the libidinal pleasures of playing god, become more disposed to recognize this.
Here, we can modify Nietzsche's ``god is dead, and we have killed him'' to ``god is dead, and it has killed itself'': contemporary machine vision comes to function under the god trick as a god when in fact it is not, and it is in this way that the notion of playing god at all might crumble. 

Why are material machines bad at playing ideal gods? Consider that phrase of Haraway's, the god trick as ``seeing everything from nowhere'', which has appeared around this essay several times. 
The phrase presents an immediate contradiction; it shouldn't make sense, like Noam Chompsky's canonical example of a syntactically valid but semantically confused phrase, ``colorless green ideas sleep furiously.''
Yet we also understand that the confusedness is part of the god trick itself; insofar as it is a false belief, the paradox illustrates this falsity.
This is worth reflecting upon, because it suggests that we can grasp meaning from a contradiction.
But contradictions are not intelligible in the ``material'' world.
An apple which falls at my feet has fallen from somewhere; there is no material sense in which an apple falls from nowhere, although the idea may have great phenomenal meaning for us (``ouch, that apple fell from nowhere!'').
The question I set forth, then, is: can machines ``see everything from nowhere''?
Machines are built from material parts and programmed to move in well-defined ways.
So, the answer is very simple: just as the apple does not fall from nowhere, a machine cannot materially see everything from nowhere.
This seems like an obvious statement, and it is, but it allows us to understand exactly how a machine which we constitute as god under the god trick is very clearly not a god, and why we might be more easily deceived by the god trick when we ourselves play the gods.
We can re-interpret the history of thought on machine vision presented earlier by Fei-Fei and Krishna as a material response to the notion of ``seeing everything from nowhere''.
To see everything, a machine must see from somewhere rather than nowhere, so researchers hand-design particular algorithms to check for particular patterns and important features in the image.
These algorithms are the ``somewhere'' from which those early object detection models looked at objects in the world.
Such models, however, failed to capture variability and dynamic nature of object categories in the world: they only saw ``somethings from somewhere''.
To truly see everything, models need to see from everywhere.
Therefore, machine vision begins about building datasets which map visual objects to seen knowledge which demonstrate achievement jointly in scale, diversity, and quality.
The interrelation between these elements should be examined closely.
At first blush, quality may appear to be in opposition to diversity.
Quality is often associated with consistency, whereas diversity works directly against this sort of homogeneity.
Why are they presented together? -- If the goal is to better map the world, and truths in the world just are jointly formed from diverse points of access, then quality and diversity are aligned.
Likewise, scale and diversity are in alignment with one another, because if the scale of collected truths-of-the-world expands, it will be diverse because these truths are diverse; the diversity of the world requires scale to capture.
To summarize these relations in Haraway's succinct language: objectivity is formed from partial perspectives.
The development of machine vision towards its self-improvement (i.e., towards the position of playing god under the god trick), then, has more or less arrived at the basic insight that Haraway directs towards science and scientific inquiry.
Hence, we can observe a large quantity of recent research in machine vision which further develops diverse partial perspectives of the world, motivated precisely by a pursuit of ``seeing everything'' -- not necessarily in a dominating sense, but towards objectivity.\footnote{For examples, see: Vikram V. Ramaswamy et al., “GeoDE: a Geographically Diverse Evaluation Dataset for Object Recognition” (NeurIPS, 2023). https://arxiv.org/abs/2301.02560. Andre Ye et al., “Cultural and Linguistic Diversity Improves Visual Representations” (arXiv, 2023). https://arxiv.org/abs/2310.14356.}

\section{Machine vision, moral vision}

Admittedly, the argument advanced so far might seems unsatisfying or at the very least ethically incomplete: while the development of machines may work towards deconstructing the god trick for objective knowledge, clearly humans can and do weaponize machines to play as very destructive gods.
As I suggested in my introduction, there are many others that have set forth human-centric analyses: critical accounts of the humans which dangerously wield machines and the social systems that produce them.
This essay, however, is an attempt to understand the ways in which machines might work, in turn, \textit{upon humans}.
It is significant that improvement of machine vision directs human labor towards mapping the world with attention towards ``scale, diversity, quality''.
I will carry forth this notion of \textit{directing labor} in an ethical analysis of machine vision. 
It may be that this kind of insight -- that machines encourage certain kinds of behavior for humans -- is the best that we can do: we cannot directly ascribe an ethical orientation (`just', `unjust', `good', `bad', etc.) to machine vision as if it really were so (that it really is just, unjust, etc.), because to do so would be another sort of god trick: to posit an objective view of ethical character while deliberately excluding the positer from the picture.
What we can do, however, is understand how machine vision directs human labor in particular ways that encourage ethical reflection.

Iris Murdoch is familiar with this notion of turning away from the seemingly ceaseless debate over the right kind of ethical judgements towards a focus on ethical labor and reflection. These, Murdoch writes, are not ``preliminaries to choice''\autocite[41]{Hepburn1956}  -- as they are treated by those ethicists who study only what choices ethical agents should make -- but rather the doing of morality itself. What happens when a moral situation is ambiguous and it is not clear how to approach it?
\begin{quote}
    What is needed is not a renewed attempt to specify the facts, but a fresh vision which may be derived from a ``story'' or from some sustaining concept which is able to deal with what is obstinately obscure, and represents a "mode of understanding" of an alternative type.\autocite[51]{Hepburn1956}
\end{quote} 	
For Murdoch, this ``fresh'' moral vision requires us not only to be presented with the basic facts of the moral situation but to find its salient moral features.\autocite[308]{Blum2012}
Rephrased, moral vision is not merely about perceiving but also about cognizing.
Cognizing in a way which fails to recognize or even distracts from the salient features of the moral situation results in ``distorted vision''.\autocite[309]{Blum2012}
Persons may have distorted vision because their cognition is structured by personal, social, and cultural ``fantasy''.\autocite[316]{Blum2012}
A father may fail to see that his son can live a good life as a philosopher because he holds a personal fantasy that his son will become a businessman like him, but perhaps also because he buys into the social and cultural fantasy of male masculinity which marks a businessman as more preferable position for a man than a philosopher.
Yet, the father may believe that he sees what is best for his son better than his son does, and domineer his son's life away from philosophy towards business. 
Haraway shows us how the god trick gives us false rather than objective knowledge; Murdoch may help us understand the ethical cost: the god trick engages us in fantasies that obscure moral reality.
So, insofar as machine vision works against the god trick, it may also work against distorted vision.

Slavoj Žižek says that behind every genocide there is a poet, a figure who ``tortures'' language to articulate a rousing complex of emotions, dreams, visions.\autocite{ZizekPoetry}
One might suggest, in turn, that there is no poetry for machines: information just is its material contents, and there is no sense for the machine in which it might be more or less.
A solider may be seduced by poetry and ideology into identifying, tracking, and killing an `enemy', precisely because they do not see the salient moral features: rather, they may cognize themselves as merely carrying out the will of the nation, justice, and history.
Therefore, Hannah Arendt observes, evil may acquire a banal quality: extermination is distorted as a bureaucratic system of collecting and disposing waste.\footnote{See: Hannah Arendt, \textit{Eichmann in Jerusalem: A Report on the Banality of Evil} (1963).} 
If genocidal poetry is our hammer and we play gods, then we may ceaselessly and banally hammer, carrying out a task whose moral reality we cannot see.
But when the hammer breaks, and we become aware of our particularity, the banality of consistency is disrupted and the fantasy fractures, even if just for a moment.
It is at this point that we are opened up to engage in ethical reflection, and the salient features of moral reality make themselves present to us.
To build a machine to identify, track, and kill an `enemy', one must collect datasets of precisely what an `enemy' look like, label their heads, write the calculation for the direction and distance of the target from the machine, program the shooting, etc.
One cannot directly give machines poetry and ideology to distort its vision of reality, but rather must spell out its desired function in material terms.\footnote{Certainly, poetry and ideology can make its way into machines because machines are built by people who are susceptible to these forces. But this does not negate the fact that machines are material beings constrained by material factors.}
Machine vision remains transparent; it does not `hide itself'.
A large and quickly growing of work probes machine learning models for distorted vision, such as disparities across race and gender, not only because it works towards more just and equitable goals in the field but also because models' distorted vision must be addressed as a research goal to achieve a more objective machine vision.
That is, distorted vision is not just an ethical question but an important technical question. Insofar as contemporary machine vision is a Heideggerian hammer rather than Mjölnir, building machine vision directs our labor towards material moral reality. 

Certainly, this may not be enough. 
We may choose not to take the chance for moral reflection when the hammer breaks; we can fix the hammer and continue hammering. 
But perhaps it is asking too much for the hammer to somehow remain broken when it is being used unjustly. 
Instead, we can recognize that what the hammer can do is to \textit{be material}, and therefore to break from time to time, to make the carpenter struggle and challenge her fantasies, even if briefly.
Haraway writes: ``Vision is always question of the power to see -- and perhaps of the violence implicit in our visualizing practices. With whose blood were my eyes crafted?''\autocite[444]{Haraway1988}  Contemporary machine vision forces us to ask precisely this question, because we are the ones who must collect the blood to craft its eyes.

\textit{Read April 6th, 2024 at Forest Grove, Oregon.}

\section*{Acknowledgements}
I would like to thank Carina Fourie for her helpful discussions and feedback on the piece.

% \newpage
\printbibliography

\end{document}